\documentclass[titlepage,12pt]{utarticle}

\usepackage[pdftex]{graphicx}
\usepackage{amsmath,amssymb,amscd}
\usepackage{color}
\usepackage{latexsym}
\usepackage{ifthen}
\usepackage[]{hyperref} 

\numberwithin{equation}{section}

\newcommand{\be}{\begin{equation}}
\newcommand{\ee}{\end{equation}}
\newcommand{\bea}{\begin{eqnarray}}
\newcommand{\eea}{\end{eqnarray}}

\begin{document}
\preprint{{\tt arXiv:yymm.xxxx}\\}

\title{Muon Tomography of Ice-filled Cleft Systems \vskip 0.5cm in Steep Bedrock Permafrost: A Proposal}

\author{{\sc Matthias Ihl~}
 \thanks{a: Affiliation at the time of writing, b: Present affiliation.}
 \address{
  Department of Physics,\\
  University of Texas at Austin\\
  Austin, TX 78712, USA\\
  Email: {\tt msihl@zippy.ph.utexas.edu}
        }
 \address{
 Instituto de F{\'i}sica, \\
 Universidade Federal do Rio de Janeiro,\\
 21941-972 Rio de Janeiro, RJ, Brasil\\
 Email: {\tt msihl@if.ufrj.br}}
}

\date{\today}

\Abstract{In this note, we propose a novel application of geoparticle physics, namely using a muon tomograph to study ice-filled cleft systems in steep bedrock permafrost. This research could significantly improve our understanding of high alpine permafrost in general and climate-permafrost induced rockfall in particular.}

\maketitle
\tableofcontents

\section{Introduction and Summary}
\begin{figure}[h]
 \centering
 \includegraphics[totalheight=0.3\textheight]{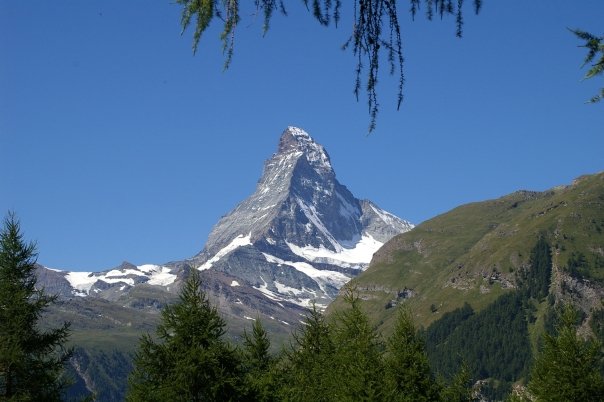}
 \caption[]{Photograph of Matterhorn. Its iconic shape closely resembles our simplified model presented below (figure~\ref{fig:setup}).}\label{fig:mh}
\end{figure}

The subject of the proposed research project is the development and application of a muon tomograph in order to study ice abundance in 
high alpine permafrost (figure~\ref{fig:mh}). To this end we want to employ suitable particle detectors to track cosmic-ray muon rates at different angles and locations
for the purpose of computer-assisted tomographic image reconstruction. Differences in count rates from different directions indicate different total masses
(density lengths) along these directions and can thus be used to create 3D tomographic images using computer-assisted tomography (CT). 
The proposed project constitutes an important new application of geoparticle physics, i.e., the application of known methods from particle physics to study geophysical problems. 
\section{State of research}
\subsection{Ice in steep bedrock permafrost}
The stability of many steep mountains is influenced by permafrost (soil or rock material that remains at or below zero degrees Celsius for at least two years). Cleft systems that are ice-filled and perennially frozen change their stability conditions when warmed \cite{DAV2001} or thawed \cite{GRU2007} and in some cases, this leads to large rock fall events \cite{gruber2004permafrost, GER2006, NOE2003}. Permafrost thus translates climate change into slope stability and affects both the evolution of mountain landscapes and natural hazard regimes. \\
While considerable progress has been made in understanding the thermal subsurface field in mountain topography \cite{GRU2004b} and in modeling its spatial patterns \cite{GRU2004c} and transient response \cite{NOE2009} to climate change \cite{NOE2007} at diverse scales \cite{RIS2008}, the distribution of ice-filled cleft systems is virtually unknown \cite{GRU2007}. \\
Currently, two candidate processes are described \cite{GRU2007} that can lead to ice-filled cleft systems and are investigated in field \cite{HAS2008, BEU2009} and laboratory studies \cite{MUR2006}. However, their relative importance and the depth below the surface to which they can act is unknown, despite the fact that their presence is the crucial ingredient for climate-permafrost-induced rockfall. Elucidating the ice content of steep alpine bedrock ridges thus is a key element in understanding the mechanisms of ice formation and, especially contributes to anticipating the potential for future rock fall from degrading permafrost areas.
\subsection{Geoparticle physics}
\begin{figure}[h]
 \centering
 \includegraphics[totalheight=0.4\textheight]{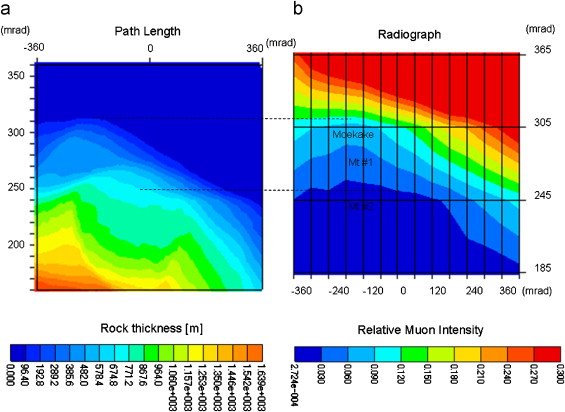}
 \caption[]{(a) Angular distribution of the path length from the topographic map in the top region of Mt. Asama. (b) The cosmic-ray muon radiograph is shown in the same angular region. Figures taken from \cite{2007NIMPA.575..489T}.}\label{fig:asama}
\end{figure}

Geoparticle physics is the science of probing geological structures by means of cosmic radiation with methods borrowed from particle physics. 
There are two subfields defined by the particle probe chosen.
An important example is neutrino radiography based on cosmic neutrinos. Several neutrino telescopes are under construction, e.g., 
IceCube under the Antarctic ice \cite{2004APh....20..507A} and the KM3NeT NEMO telescope under the mediteranean sea \cite{Katz:2006wv}.
Neutrinos also feature prominently in some of the observations made at the Pierre Auger Observatory \cite{2006APh....25...14A,anchordoqui:123008}. Cosmic neutrinos are useful when one wants 
to probe deep below the Earth's surface or explore its core and mantle, due to their extemely large interaction lengths. Therefore they are a unique experimental tool to study the Earth's density profile.  
In contrast to cosmic neutrinos, atmospheric muons, i.e., muons produced by the interaction of cosmic rays with the atmosphere, can only penetrate through a few 
kilometers of rock at most, thus they can only be used to probe superficial structures such as the internal structure of the Earth's crust or mountains. 
However, to this day, superficial structures are most commonly studied by electromagnetic, gravitational or seismologic geophysical techniques which are indirect measurements with large intrinsic uncertainties. Muon radiographic or tomographic methods present an important alternative tool to study superficial structures such as steep rock permafrost. These methods most importantly allow to measure the density of the structure, a prominent property that can be readily interpreted in terms of physical composition and state.
In addition, muon radiography can provide information about the chemical composition and thermal state.
Especially at sites where the subsurface structure cannot be well resolved by conventional techniques due to its inaccessability or heterogeneity, muon detectors are suitable alternatives to provide a map of density variations.\\
Muon radiography was first proposed by L.W. Alvarez and collaborators in 1970 and employed to 
potentially discover unknown cavities inside the Second Chephren pyramid \cite{1970Sci...167..832A}. A similar program is underway in the group of Prof. Schwitters at the University of Texas at Austin, where a muon detector 
is being built for the purpose of exploring the internal structure of Mayan pyramids in Belize \cite{UTMayaMuon2004}. Groundbreaking work in muon radiography has been done recently by H. Tanaka and collaborators in Japan \cite{2007NIMPA.575..489T, 2007E&PSL.263..104T, 2007GeoRL..3422311T} . Their observations to determine the density distribution of the summit crater of Mt. Asama clearly demonstrate the potential of techniques using cosmic-ray muons (figure~\ref{fig:asama}).
\subsection{Detector designs}
There are two different types of detectors that can be used to track muon trajectories and count rates
for the purpose at hand. The mountain environment that the detectors will be operated in necessitates special requirements like portability, low or zero electrical power consumption and minimal need for maintenance. A full detector setup will consist of many detector planes to form a so-called "telescope".\\ 
 The detectors serve three primary purposes which are summarized below:
\begin{itemize}
\item Detection of single cosmic-ray muons (events) to enable counting rates at which muons hit the
target volume. Comparison of rates from different muon directions provides information about 
absorption rates and thus about the total mass along those muon ray paths.
\item Precise measurement of each muon's trajectory to allow a meaningful extrapolation of  the
muon's path through the object to the target volume (which can be assumed to be pointlike compared to the object under study). 
\item Determination of the muon's energy (velocity) to reject low-energy muons whose large scattering would not allow to extrapolate the trajectory to a sufficient degree of accuracy. Alternatively, with certain detector types, it is possible to incorporate a threshold which rejects muons whose velocity is below a certain value. 
\end{itemize} 
In the following I describe and compare the advantages and disadvantages of the different designs.
\subsubsection{Scintillation detectors (SDs)}
The most common technique for muon detection used in geophysics to date is based on plastic scintillators strips with 
wavelength shifting (WLS) fibers frequently used in high energy experiments \cite{2001NIMPA.466..482P} such as MINOS at
Fermi lab \cite{2006NIMPA.556..119A} or OPERA at the Gran Sasso National Laboratory \cite{2004PAN....67.1092D}. 
Polysterene scintillator strips are ionization/location detectors that track ionization signals which are subsequentally converted into data appropriate for computer analysis by employing WLS fibers,
photomultiplier tubes (PMT) and front-end electronics. The WLS fibers are located in a small groove on the scintillator surface. The scinillation light emitted when ionizing particles pass through
the scintillator are absorbed by the WLS fibers, re-emitted at a slightly different wavelength and transmitted to the PMT tubes at both ends with insignificant losses. \\
The traditional rectangular design has advantages in being simple, robust, relatively inexpensive and well understood by the MINOS group. However, more modern geometries, featuring triangular sections, significantly improve the spatial resolution compared to rectangular scintillator designs. With different layering techniques, total sizes of scintillation detectors vary from a few $m^2$ to thousands of $m^2$ in MINOS and OPERA.\\
In order to measure the muon velocity, the UT Muon Maya Group employs a Cherenkov detector.
This is important to be able to reject low energy muons which feature large scattering inside the object under scrutiny, thus deteriorating the theoretical resolution significantly (cf. below).
However, the size, cost and weight exceed acceptable limits for our type of application. However, there have been significant achievements in reducing size and weight of this type of detectors recently. With further improvements, this technology may become a feasible option in the future.\\ 
The advantages and disavantages of the scintillator design are listed below.\\
Advantages:
\begin{itemize}
\item SDs are commonly used in a wide variety of applications, therefore they are relatively inexpensive in production and well understood.
\item Large detector areas can be achieved at low cost.
\item With moderate detector areas, the power consumption is low enough to be feasible in remote 
environments. The prototype detector of the Vesuvius group in Italy \cite{MURAY2008} uses approx. 70 W of power which can be produced on site, e.g., with solar panels.
\item SDs are extremely robust and insensitive to environmental condtions. Thus they can be operated in extreme conditions with minimal on site maintenance.
\item  Data acquisition and analysis are very fast. They constitute well studied problems for which software packages are readily available.
\end{itemize}
Disadvantages:
\begin{itemize}
\item Preliminary estimations of the necessary detector area indicate that, for the purpose at hand,
the total area should be on the order of magnitude of 100 $m^2$ to achieve reasonable exposure times.
While the production of such a detector is possible within reasonable  bounds of effort and cost, the size makes it difficult to transport and install such a system at a remote mountain site.
\item The main disadvantage of SDs compared to the emulsion detectors discussed below is their spatial resolution. With the aforementioned setups, angular resolutions of down to 10 to 15 mrad in  optimal detection conditions seem plausible. This corresponds to a spatial resolution of 1 to 1.5 meters at 100 m distance.
This is roughly a factor 10 worse than what can be achieved with nuclear emulsion detectors.    
\end{itemize}

\subsubsection{Nuclear emulsion detectors (EDs)}
Detectors using nuclear emulsion films represent a very old observational technique (originally discovered by H. Becquerel, the phenomenon was recognized as being related to the ionization of charged particles as early as 1909) in particle physics. Recently, there has been renewed interest
sparked by technological advances made in Japan during the past few years \cite{2007NIMPA.575..489T, 2007E&PSL.263..104T, 2007GeoRL..3422311T}.\\
Most importantly, the production technology of the emulsion films (optimization of gel properties) and photographic techniques were improved in such a way that 
the nuclear emulsion technique now is feasible for large-scale applications, such as OPERA \cite{2006NIMPA.556...80N}.
Secondly, significant developments in automated emulsion read-out systems nowadays allow for 
a speedy data analysis that proved to be extremely difficult before. The read-out speed is envisaged to reach 
roughly 10 $m^2$ of film per day per lab in the near future. Such a system consists of an automated read-out system,
e.g., the Ultra Track Selector (UTS), and off-line software for further processing of the huge amount of read-out data. Another important new feature is the capability of emulsion-refreshing, i.e., erasing unwanted tracks before use by controlling the fading properties of the gel.
A detailled account of these improvements can be found in \cite{Kodama:2000mp, Kodama:2002dk}.\\
Modern nuclear emulsion cloud chambers comprise layers of emulsion film alternating with lead plates to achieve a sufficiently high target mass. The emulsion film layers contain micro crystals of 
AgBr, which become developable after a charged particle passes through. Therefore a particle trajectory can be tracked after development as a sequence of three dimensionally ordered grains.
Emulsion films can contain huge amounts of data stored as photographic images. The real challenge is to convert those pictures into usable data. As mentioned above, this challenge has been successfully adressed in recent years. \\
Advantages:
\begin{itemize}
\item Highest resolutions of down to 1 mrad can be achieved which correspond to a spatial resolution of 10 cm at a 100 m distance. As with SDs, the resolution is naturally limited by muon scattering. However, the lead plate layers
can effectively be used to exclude low-energy muons, thus further improving resolution.
\item EDs are passive devices, i.e., zero power consumption, extremely suitable for the remote environment they are supposed to operate in. 
\item The size and weight of the EDs are minimal, thus it will be easy to carry them to and install them at their intended locations.
\item Due to their small size it is relatively easy to effectively protect the EDs against the harsh environmental influences at remote high alpine mountain sites. Moreover they are expected to be fully functional in a wide range of temperatures as long as the interlayer temperature gradient is not too large, which is not expected to be problematic.
\item Lastly, the costs for production and operation are minimal as well.  
\end{itemize}

Disadvantages:
\begin{itemize}
\item The read-out and data recovery speed is much slower than conventional systems based on PMTs and front-end electronics. However, as mentioned above, there will be significant improvments in the foreseeable future. Moreover, the estimated measurement times are on the order of weeks to months, therefore data processing times of days to weeks are still acceptable.
As part of the OPERA collaboration, LHEP at the University of Bern is equipped with the most modern automated emulsion read-out microscopes in Europe. Even more capacities will become available shortly.  
\end{itemize}
In conclusion, the nuclear emulsion detectors seem to be the superior choice of detection device for the purpose at hand. In the following we will therefore develop an outline for a research program based on this detector technology.

\section{Scientific goals}
The eventual goal of the proposed project is to produce 3D images of high alpine permafrost regions inside mountains, using a muon tomograph as described below, in order to determine and map their geological composition, with an emphasis on ice content. The importance of this was outlined in section (3.1).
To this end, I want to formulate the following research questions that will serve as a guideline in formulating intermediate goals for the project. They will be elaborated further in the next section.
\newpage
\noindent{\bf Research questions}:\\
From a geoscience point-of-view:
\begin{itemize}
 \item What are the abundance and distribution of ice-filled cleft systems in permafrost of steep alpine bedrock ridges?  
 \item How does climate change influence the slope stability and evolution of mountain landscapes?
 \item Can we establish a model of climate-permafrost-induced rock fall in order to identify potential future rock fall from degrading permafrost?
\end{itemize}
\noindent From a physics point-of-view:
\begin{itemize}
 \item Can the technology of muon tomography (that is well studied and understood in the realm of particle physics) be successfully adapted and applied to a geophysical setting such as the one proposed herein (with a view toward the difficult environmental conditions)? 
 \item What is the best resolution that can be achieved with nuclear emulsion detectors in this framework and what is the maximal thickness of rock that will be accessible under realistic conditions?
 \item What are the exposure/measurement times necessary to gather reliable data in a high alpine setting?
\end{itemize}

\noindent{\bf Intermediate goals}:
\begin{itemize}
 \item Develop and build a fully functional prototype detector according to specifications and durability requirements in order to enable equipment testing 
 under realistic conditions and to measure muon fluxes experimentally on site. 
 \item Identify a suitable mountain ridge for which sufficient topographical data is readily available or can be easily obtained to ensure
 suitable (i.e. small enough) potential angular resolutions.
 \item Optimize the detector-mountain setup using experimental muon flux and transmission data, and Monte Carlo simulations.
\item Development of a computer-assisted tomography software that reconstructs 3D images from the data taken by the system of detectors.
\end{itemize}

\section{Research program}
\begin{figure}[h]
 \centering
 \includegraphics[totalheight=0.25\textheight]{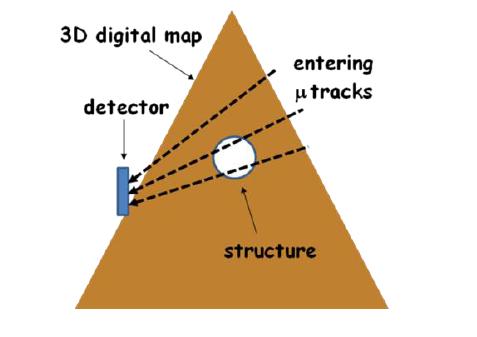}
 \caption[]{Sketch of the mountain-detector setup. The depicted idealized internal structure is assumed to be spherical (despite the fact that cleft systems will be planar features).}\label{fig:setup}
\end{figure}

The main task of the research program is to study ice abundances in steep bedrock permafrost (with a view towards more 
general applications of studying the internal structure of mountains). To achieve this, we 
propose to employ a suitable emulsion detector system to track cosmic-ray muon rates at different angles (and different
locations) for the purpose of 3D tomographic image reconstruction. This is possible because differences in count rates from different angles
indicate different density lengths along those directions which allows to detect internal structures (such as ice-filled cleft systems with a typical density of almost 1 $g$ $cm^{-3}$) embedded in the surrounding rock (with an assumed standard density of 2.5 $g$ $cm^{-3}$) and enables computer-assisted tomographic reconstruction.
A simplified model sketch of the setup is drawn in figure~\ref{fig:setup}.

\begin{figure}[h]
\centering
\includegraphics[totalheight=0.25\textheight]{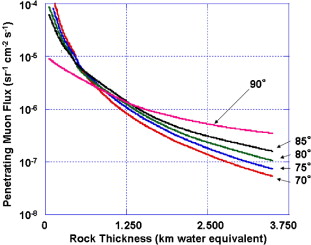}
\caption[]{Integrated flux of cosmic ray muons at various zenith angles pentrating through a given thickness of rock with a density of $2.5 g/cm^3$. }\label{fig:flux1}
\end{figure}

The research program has to be structured and devided into several main stages of research and development. \\
At first it will be necessary to identify a suitable object of study, i.e., a potentially interesting mountain ridge
or steep rock face that fulfills our requirements concerning size and shape. Its size should not exceed a few to several hundred meters of rock. Moreover, we need sufficient topographical data
at an accuracy of $\leq 5$ cm to be able to achieve precision measurements in the vicinity of 1 mrad
angular resolution. The technique presented herein is especially suitable for thin rock layers, i.e., little rock thickness. The reason is that, for muon energies above 100 GeV,  
the muon flux has a power law dependence on the energy, with exponent -2.7, and thus falls off rather steeply. Therefore the observation of muons traversing thick layers of rock becomes increasingly difficult. The energy loss of relativistic particles in matter can be estimated to be roughly 
2.5 MeV $g^{-1}$ $cm^2$. Assuming a standard rock density of 2.5 g $cm^{-3}$, the energy loss is a little more than 0.5 TeV $km^{-1}$. The integrated muon flux as a function of rock thickness is plotted in figure~\ref{fig:flux1} (taken from \cite{2007E&PSL.263..104T}.)

Furthermore, it will be necessary to adapt the nuclear emulsion detectors to our purposes, most importantly to install them in a weatherproof casing and to make sure they can withstand the extreme temperature differences and environmental conditions on site.
The next step is to determine the optimal specifications and locations for the detectors. This includes a simulation of the experimental apparatus and estimation of muon fluxes. To this end, one has to do feasibility studies starting with a Monte Carlo simulation to estimate the various muon fluxes and transmission rates. The optimization process comprises gradual changes of detector design and locations and will lead to the final optimal design and locations of the emulsion detectors.
For the Monte Carlo simulations, a wide variety of software packages (e.g., GEANT4 \cite{Agostinelli:2002hh}) is readily available and has been tested in many (sometimes similar) applications. A digital elevation map of the mountain
will be the geometrical input to the Monte Carlo simulation. In a first step, under some hypothetical
assumption of the internal structure, one generates a large number of tracks which run backwards from the detector and intersect the digital elevation map and the hypothetical substructure to be 'revealed'. For each tracjectory, the different layers of matter and their density are recorded. This data subsequently serves as input for GEANT4 with the aim of determining muon transmission fluxes. This should be supplemented by actual experimental muon measurements
on site, helping to understand the experimental problems and to validate the Monte Carlo simulations. At the end of this stage one should be able to produce first radiographical images that could be compared to conventional geophysical techniques.\\
Thirdly, simplified model calculations of the mountain-detector system are necessary to estimate the detector size (effective area) and exposure time. A rough estimate based on the simplified model depicted in figure~\ref{fig:setup} suggests that exposure times on the order of a few weeks are realistic. This will likely be improved once the exact detector configuration has been determined and allows for realistic simulations and testing.\\
Ultimately, the goal is to install several detectors at optimized locations to be able to reconstruct
3D tomographical images of the internal structures. This final step will take a strong effort of software development but, fortunately, tomographic methods are widely used in medical and particle physics and are well studied and documented.  

\section*{Acknowledgements}
I am indebted to Stefan Gruber for suggesting this project, numerous discussions and proofreading an earlier version of this note. I would also like to thank Antonio Ereditato, Sonia Paban and Roy Schwitters for discussions and comments.

\end{document}